\documentclass[tightenlines,twocolumn,prl,nofootinbib]{revtex4-1}

\usepackage{graphicx}
\usepackage{subfigure}
\usepackage{bm}
\usepackage{multirow}
\usepackage{amssymb}
\usepackage{hyperref}

\begin{document}

\pagestyle{plain}


\title{Antineutrino monitoring for the Iranian heavy water reactor}

\author{Eric Christensen}

\author{Patrick Huber}
\email{pahuber@vt.edu}

\author{Patrick Jaffke}

\affiliation{Center for Neutrino Physics,
  Virginia Tech, Blacksburg, VA 24061, USA}

\author{Thomas E. Shea}
\affiliation{TomSheaNuclear Consulting Services, G\"orgengasse 10/25, 1190 Vienna, Austria}

\date{\today}

\begin{abstract}
In this note we discuss the potential application of antineutrino
monitoring to the Iranian heavy water reactor at Arak, the IR-40, as a
non-proliferation measure. We demonstrate that an above ground
detector positioned right outside the IR-40 reactor building could
meet and in some cases significantly exceed the verification goals
identified by IAEA for plutonium production or diversion from declared
inventories. In addition to monitoring the reactor during operation,
observing antineutrino emissions from long-lived fission products
could also allow monitoring the reactor when it is
shutdown. Antineutrino monitoring could also be used to distinguish
different levels of fuel enrichment. Most importantly, these
capabilities would not require a complete reactor operational history
and could provide a means to re-establish continuity of knowledge in
safeguards conclusions should this become necessary.
\end{abstract} 

\maketitle

\begin{figure}
\includegraphics[width=\columnwidth]{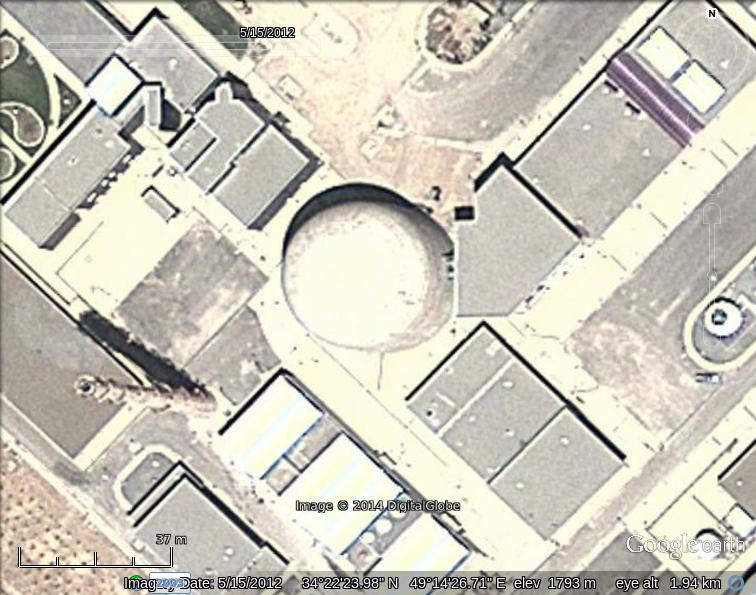}
\caption{\label{fig:position} Satellite image of the heavy water
  reactor at Arak, Iran, May 2012. Image credit Digital Globe and
  Google Earth.}
\end{figure}

The IR-40 reactor in Iran is of particular concern, since the design
thermal power of $40\,\mathrm{MW}_\mathrm{th}$ combined with the
choice of moderator, heavy water, makes this reactor ideal for
plutonium production for nuclear weapons~\cite{Heinonen:2011}; a
satellite image of the Arak reactor complex is shown in
Fig.~\ref{fig:position}. Iran states that this reactor will be used
for the peaceful purposes of isotope production for medical uses and
scientific research. It remains to be seen whether Iran will operate
the reactor at all and, if the IR-40 becomes operational, whether it
will operate as designed or with some modifications that make it less
amenable to weapon plutonium production~\cite{Heinonen:2011}, or
whether an extra-territorial siting arrangement might allay
proliferation concerns~\cite{Shea:2009}.

If the IR-40 goes into operation, the IAEA will need to confirm that
its operations are as declared, using a combination of methods that
are reliable and cost-effective. Antineutrino monitoring could
complement other methods and provide important additional assurance to
the international community that Iran continues to honor its
commitments. Existing safeguards methods are ill-suited to deal with
possible break-out scenarios or situations when inspector access is
intermittent. The historic example of the Democratic People's Republic
of Korea (the DPRK) and its interactions with the IAEA and the
international community from 1992-1994 included both
intermittent denials of inspector access and the DPRK's eventual break-out
from the NPT. As a result, the question of plutonium production in the
DPRK prior to 1994 is still unresolved, see for instance
Ref.~\cite{GoingCritical}.

Antineutrino monitoring was first proposed more than 30 years
ago~\cite{Borovoi:1978} and is based on the fact that the number of
antineutrinos produced and their energy spectrum depends in a
well-defined manner on the reactor power and on the relative
contribution to fission from the various fissile isotopes:
uranium-235, plutonium-239, uranium-238 and plutonium-241. In a recent
analysis ~\cite{Christensen:2013eza} we were able to show that the
application of antineutrino monitoring would have been able to provide
timely information about plutonium production in the DPRK -- even
given the actual, constrained and intermittent access by IAEA
inspectors. We have applied the techniques developed in
Ref.~\cite{Christensen:2013eza} to the specific case of the Arak IR-40
reactor in Iran to show that antineutrino detectors could provide the
IAEA with a resilient high-level monitoring capability not offered by
any other known technique.

The IR-40 is capable of producing 10\,kg of weapon-usable plutonium
per year. A safeguards regime for the IR-40 must be able to verify
that the actual plutonium production agrees with the declarations made
by Iran, and that the plutonium produced remains accounted
for. Obtaining plutonium from most reactors and in particular from the
IR-40, requires the reactor to be shutdown for the irradiated fuel
to be removed. To quantitatively address the diversion problem
involving plutonium from a known reactor, two questions have to be
distinguished: the total amount of plutonium produced in the reactor
and the amount of plutonium actually residing in the reactor
core. The former can be inferred from the complete power history of
the reactor, whereas the latter requires additional detailed
information on the fueling history of the reactor or a method to
directly assess the core state in terms of average fuel burn-up. It is
the agreement or disagreement of these two quantities, the total
produced and actual core plutonium, which may indicate whether or not
a plutonium \emph{diversion} has taken place.

The power history of a reactor\footnote{Total integrated reactor power
  can also be used to estimate the production of tritium in the heavy
  water inventory.} can be inferred by measuring the primary coolant
flow rate and temperature drop using a thermo-hydraulic monitoring
system, a method the IAEA already employs in some research
reactors~\cite{thermo}. The core burn-up is not usually measured
directly but is inferred from knowing the type of fuel that goes into
the reactor core and on a burn-up calculation based on the power
history of the reactor. For discharged fuel typically only the fact
that individual fuel elements emit intense ionizing radiation is
verified using Cerenkov light. The key to the relatively high
reliability of this chain of inferences is to maintain continuity of
knowledge by employing containment seals and surveillance measures.

Once continuity of knowledge is lost, recovery is difficult and
may be limited.  More sensitive monitoring methods are available to
detect complex removal scenarios, although these methods are seldom
used because they require isolating individual fuel elements, require
lengthy measurement periods, and are expensive to employ. Antineutrino
monitoring could provide a robust and non-intrusive alternative method
to recover from a loss of the continuity of knowledge.

Consider a hypothetical IR-40 example inspired by the historic DPRK
record: assume that there has been full safeguards access for N-1
months but, in the N$^\mathrm{th}$ month, continuity of knowledge is
lost.  Assume further that the reactor is shut down at the beginning
of the N$^\mathrm{th}$ month.  There could be many reasons for such
events to happen, spanning the gamut from legitimate operational
reasons, to a mere technical glitch over a diplomatic stand-off, to an
attempt at proliferation with a wide range of measures taken to delay
detection and reprisal~\footnote{For an extended period without
  inspector access, secondary means of monitoring reactor operation,
  e.g., infrared satellite imaging, could detect reactor operation and
  provide a rough estimate of reactor power.}. In the N$^\mathrm{th}$
month of our hypothetical IR-40 scenario, the power history is
interrupted, but for a sufficiently short time such that the extra
burn-up that could be achieved is very limited and therefore, does
not play a major role.  But did a refueling take place? --  The basic
task is to reestablish verifiable knowledge of the core state without
being able to rely on a power record or uninterrupted containment and
surveillance.

In Ref.~\cite{Christensen:2013eza} we showed that measuring the
composite energy spectrum of antineutrinos emitted from a reactor
could allow the burn-up and, thus, the plutonium content to be
estimated accurately and in a timely manner. We make the same
assumptions here about the detection system as in
Ref.~\cite{Christensen:2013eza}, i.e. $4.3\times10^{29}$ target
protons at a hypothetical efficiency of 100\%, which translates to a
detector mass of 10-15\,t, once the actual efficiency and chemical
composition are accounted for. We envisage a system where the whole
detector with supporting electronics fits inside a standard 20$'$
shipping container. Smaller detectors would also work but the times
required to achieve the performance we cite would be correspondingly
longer.  Furthermore, we assume sufficient background rejection
capabilities to allow for surface deployment. From
Fig.~\ref{fig:position} we estimate the diameter of the IR-40 reactor
containment building to be approximately 34\,m and therefore with the
shipping container positioned right against the exterior of the
reactor containment building, the antineutrino detector would be
located 17.5\,m from the center of the reactor core\footnote{More
  precise distances could be obtained during design information
  verification activities at the IR-40.}.

Assuming the reactor is running at full power when inspector access is
resumed, following the methods given in
Ref.~\cite{Christensen:2013eza}, the antineutrino emissions could be
used to determine the core plutonium content and, thus, to also
determine whether or not the reactor had been refueled during the
period when the inspectors were not allowed access. This burn-up based
analysis relies on standard reactor physics calculations made using
commercially available software\footnote{We carried out a reactor
  simulation of the IR-40 using the two dimensional neutron transport
  analysis code NEWT and the depletion code Origen.  Both codes are
  from the SCALE software suite~\cite{scale}.}. It provides a means to
correlate the fission rates of the various fissile isotopes in the
reactor core.  For our hypothetical IR-40 example, we assumed that the
core in its original configuration contained 10\,t of natural uranium
dioxide, and that the reactor ran at its design power of
40\,MW$_\mathrm{th}$.  Our model was derived from a full three
dimensional analysis developed by Willig, {\it
  et. al.}~\cite{Willig:2012}. Our results in terms of isotopic
composition for the major fissile isotopes and all of the main
plutonium isotopes agree to within 1-2\% with the corresponding values
reported by Willig.

\begin{figure}[t]
\includegraphics[width=\columnwidth]{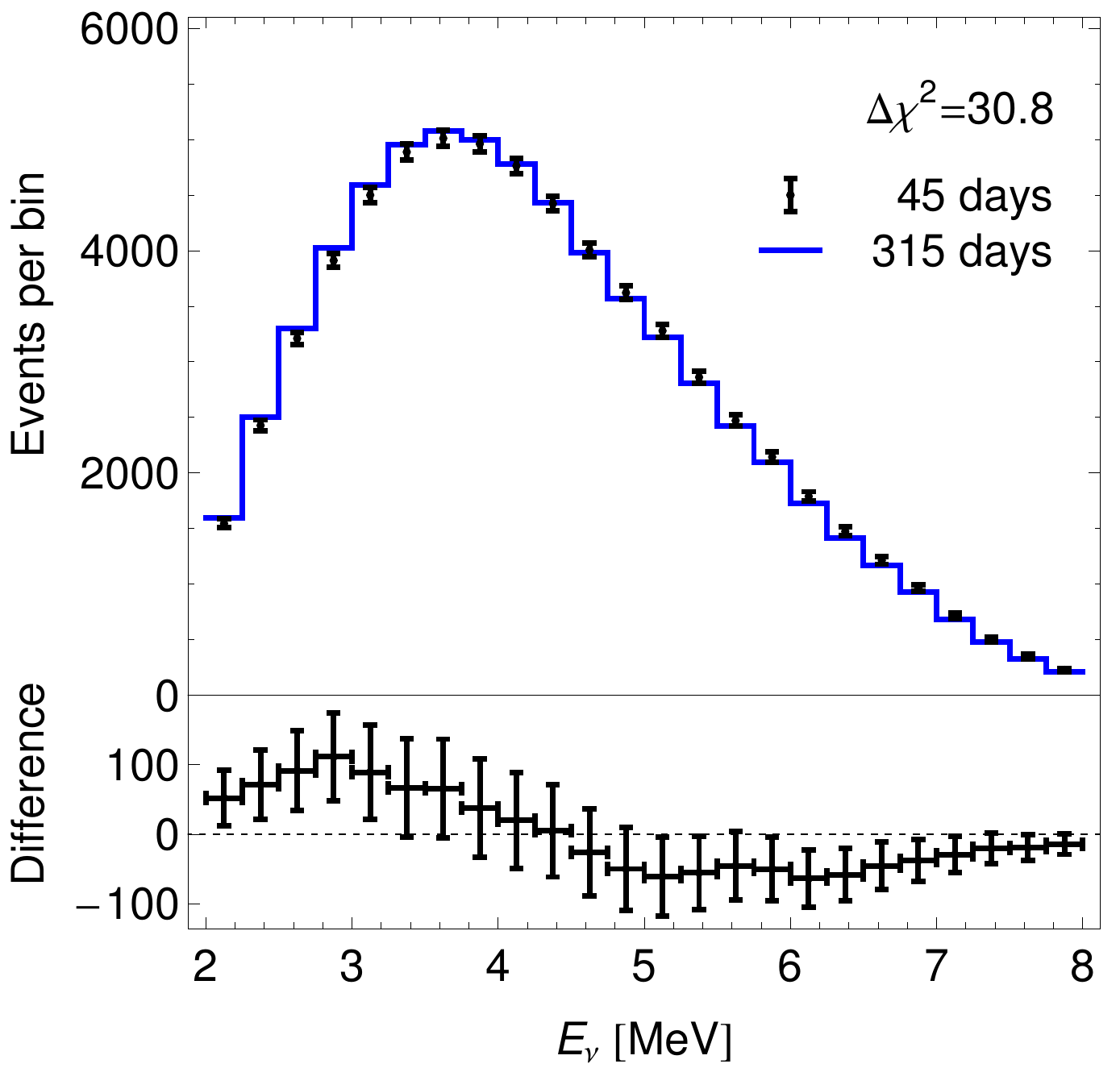}
\caption{\label{fig:rates} In the upper panel, data points show the
  event rate spectrum obtained in a 90 day data taking period for a
  core of average age of 45 days. The error bars indicate the
  statistical error in each bin. The blue line indicates the
  corresponding expected event rate spectrum for a core of average age
  of 315 days. The lower panel shows the difference in event rates
  between the 45 day core and the 315 day core and the corresponding
  statistical error bars.}
\end{figure}
In Fig.~\ref{fig:rates} we show the resulting event rate spectrum for
a core of 45 day average age (data points with statistical error bars)
and for comparison the expected event rates for a core of 315 days of
age (blue line). Clearly, the older core has a much softer
antineutrino spectrum, which is because of the much higher plutonium
content as fission of plutonium produces a softer antineutrino
spectrum. The difference in $\chi^2$ between the two cores is 30.8
units corresponding to about 7\,kg difference in plutonium
content. The visibility of this effect does not rely on extremely good
energy resolution since the spectral feature is essential bi-modal:
below about 4\,MeV the rate goes up and above it goes down.

\begin{figure}[t]
\includegraphics[width=\columnwidth]{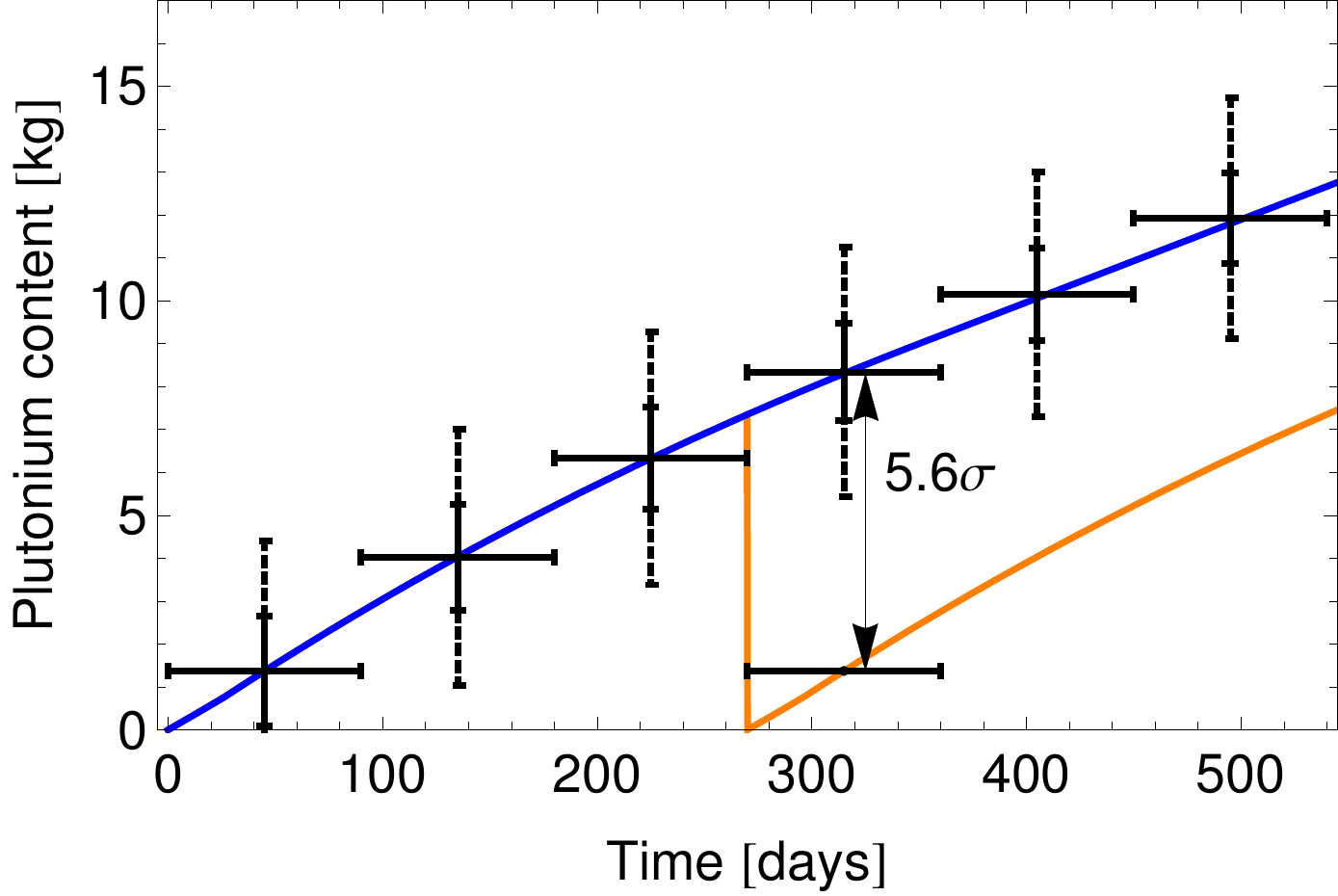}
\caption{\label{fig:pu} Shown is the 1$\,\sigma$ accuracy for the
  determination of the plutonium content of the reactor as a function
  of time in the reactor cycle. The data taking period is 90 days
  each. Dashed error bars indicate the accuracy from a fit to the
  plutonium fission rate $f_\mathrm{Pu}$, whereas the solid error bars
  show the result of a fit constrained by a burn-up model. The blue
  line indicates operation without refueling and the orange line indicates
  operation with a refueling after 270 days.}
\end{figure}

The quantitative results of our IR-40 analysis in terms of plutonium
content are shown in Fig.~\ref{fig:pu}, where the vertical axis shows
the amount of plutonium in the reactor core as a function of time. The
blue curve shows the evolution of plutonium content assuming that no
undeclared refueling has taken place, whereas the orange curve assumes
that the previously irradiated core, containing 8\,kg of plutonium,
was replaced with a fresh core after 270 days of irradation. Here, 270
days was chosen since according to Willig {\it et al.} the content of
plutonium-239 drops to 93\% after 270 days and thus 270 days
represents the longest operational period that still yields
weapon-grade plutonium\footnote{Even lower grade plutonium can be (and
  has been) used to make nuclear explosives and 93\% does not
  constitute a sharp boundary.}.  Within the first 90 days of the
putative IR-40 shutdown the two cases would be distinguished
unequivocally by analyzing the antineutrino monitoring data. Even
partial core refuelings corresponding to as little as 1.9\,kg of
removed plutonium could be detected at 90\% confidence
level. Alternatively, a full core refueling would be detected within
about 7 days at 90\% confidence level.

If the IR-40 remains shut down after the loss of continuity of
knowledge, the antineutrino detector still offers a method to assess
the core state by measuring the antineutrino emissions from the
long-lived fission fragment isotopes: strontium-90 with a half-life of
28.9\,y, ruthenium-106 with a half-life of 372\,d, and cerium-144 with
a half-life of 285\,d. In the decay chains of these three isotopes, 
antineutrinos are emitted with sufficient energy to be detected by a
standard antineutrino detector using inverse beta-decay. These
long-lived fission fragment isotopes have direct fission yields in the
percent range and thus their abundance is large and directly
proportional to the burn-up of the fuel. By measuring these
antineutrino emissions it could be possible to assess the approximate
fuel burn-up and plutonium content, and to determine whether a major
removal of spent fuel had taken place.

\begin{figure}[t]
\includegraphics[width=\columnwidth]{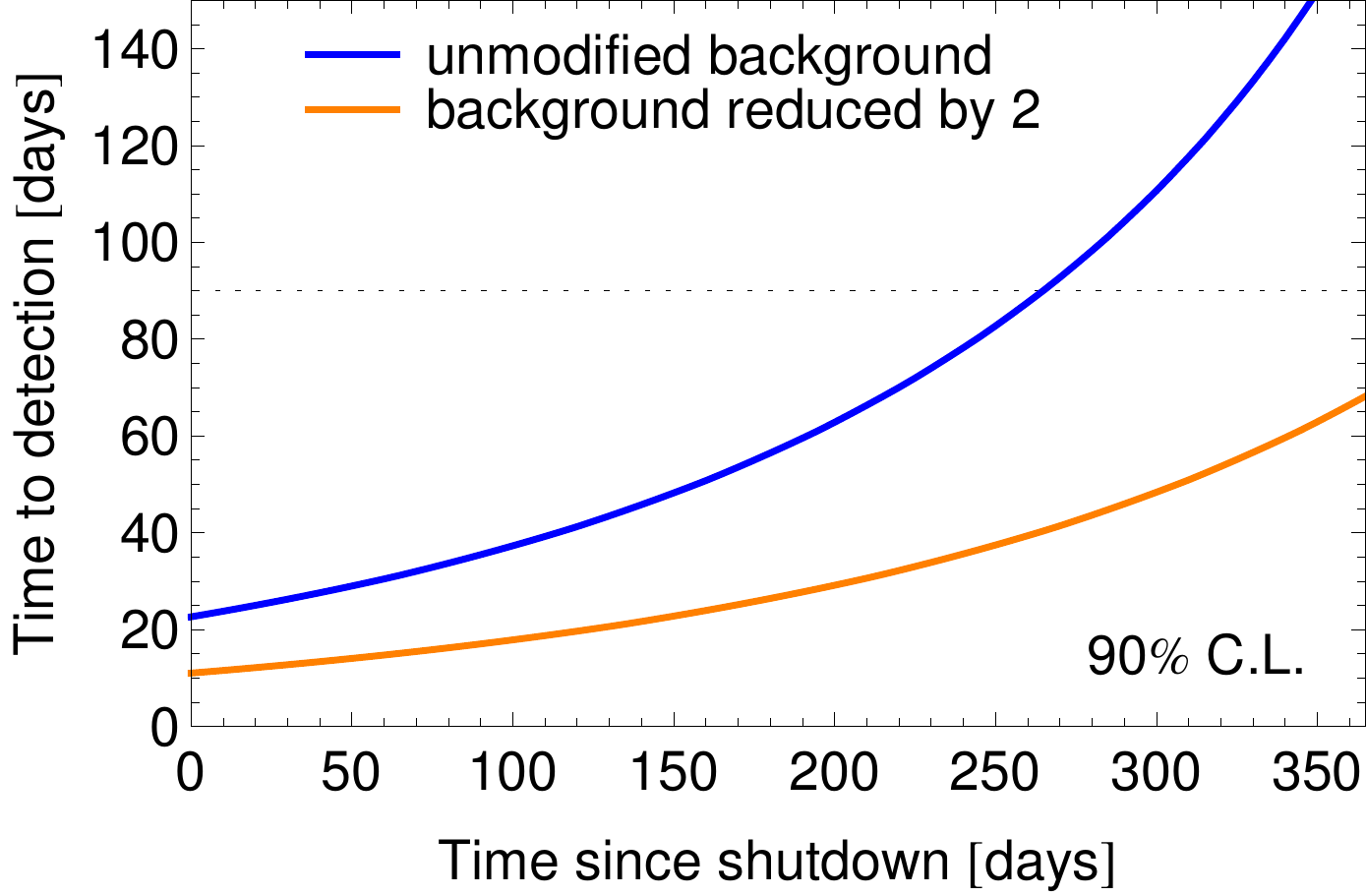}
\caption{\label{fig:lli} Shown is the time required to achieve a 90\%
  C.L. detection of defueling the reactor as a function of the time
  since shutdown when the defueling takes place. This calculation
  assumes a fuel burn-up corresponding to 270\,d of reactor operation
  at nominal power of 40\,MW$_\mathrm{th}$. The different lines are
  for different levels of cosmogenic background suppression.}
\end{figure}

The measured antineutrino rates from these fission products would be
much smaller than the antineutrino measurement rates during reactor
operation. In Ref.~\cite{Christensen:2013eza} we estimated (based on
data from~\cite{Abe:2012ar}) that there will be about 43 background
events per day per tonne of detector from beta-delayed neutron
emission from cosmogenically produced lithium-9 and about 1 background
event per day per tonne from fast neutrons. In Fig.~\ref{fig:lli} we show
the time required to achieve a 90\% confidence level detection of
removal of all the spent fuel contained in the reactor core as a
function of the time since shutdown when the core removal occurs.

As previously stated, the size of the signal is proportional to the
burn-up of the spent fuel, hence the longer the reactor has been
running the easier this measurement becomes. Even in the low
burn-up case, which would be characteristic for the production of
weapon-grade plutonium, this measurement could be performed with
current background rates for data taking as late as 250 days after
shutdown. Or, in other words, if data taking starts within a month
after shutdown, a 90\% confidence level confirmation of the presence
of the core can be achieved within 30 days or less. With a moderate
improvement in background rejection by a factor of approximately 2,
this measurement could succeed even a year after the shutdown.

Given the proliferation concerns regarding the IR-40, it has been
suggested that the reactor could be modified to make it less suitable
for the production of weapon-grade plutonium. One possibility would be
to modify the reactor to use low-enriched uranium (LEU) instead of
natural uranium (NU) as a fuel and changing the moderator from heavy
to light water~\cite{Heinonen:2011}. A detailed neutron transport
reactor physics calculation has been reported by Willig {\it et
  al.}~\cite{Willig:2012}.  They concluded that changing the moderator
from heavy to light water could be detrimental to reactor
safety. Instead it has been proposed to use a heavy water moderator
together with fuel enriched to 3\%, providing a use for the
existing Iranian stock of LEU. This LEU configuration could reduce the
annual plutonium production from 10\,kg to 3.9\,kg with a slightly
smaller fraction of plutonium-239.

\begin{figure}[t]
\includegraphics[width=\columnwidth]{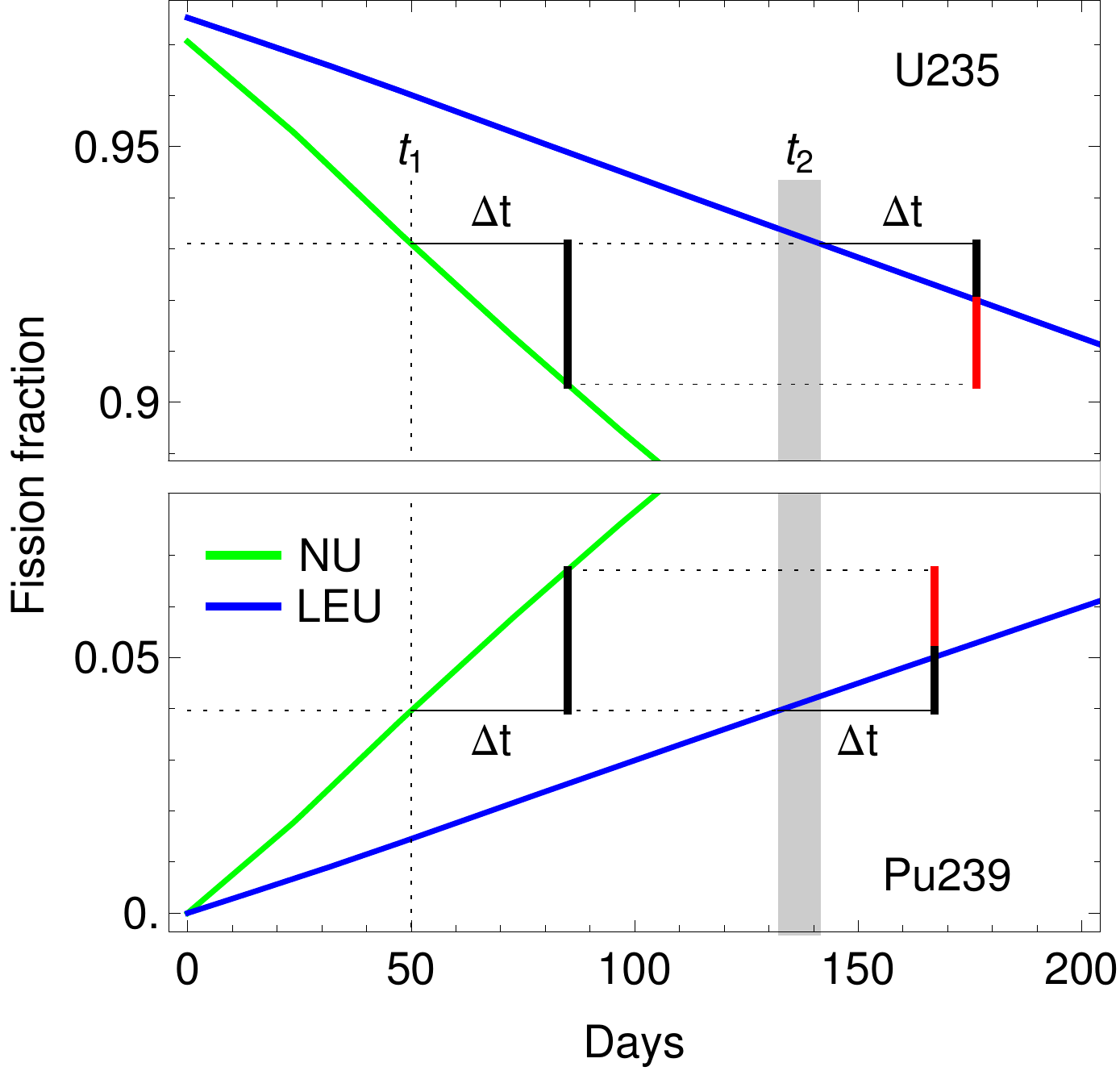}
\caption{\label{fig:enrichment} Shown are the fission fractions in
  uranium-235, upper panel, and plutonium-239, lower panel, for a
  natural uranium fueled core (NU) in green and for a 3\% enriched
  uranium fueled core (LEU) in blue as a function of time elapsed in
  the reactor cycle.  The fission fractions in both isotopes at time
  $t_1$ for the NU core match those at a later time $t_2$ for the
  LEU core as indicated by the horizontal dashed lines. The change of
  fission fractions after a fixed time interval, $\Delta t$, the so
  called differential burn-up, is indicated by the thick vertical
  black lines. There is a distinct difference in differential burn-up
  between the LEU and NU cores for both isotopes as indicated by the
  thick red lines.}
\end{figure}
If LEU fuel were introduced into the IR-40, antineutrino emissions
could also be used to distinguish a natural uranium fuel core from a
low-enriched uranium configuration by tracking the rate of change in
the plutonium fission fractions in the reactor, a technique we term
{\it differential burn-up analysis} (DBA). The basic observation
behind DBA is that both configurations follow the same overall
burn-up pattern: specifically, for the uranium-235 fission fraction,
$\digamma_{U235}$, and the plutonium-239 fission fraction,
$\digamma_{Pu239}$.  Being on the same overall path implies that
looking at a single snapshot in time, $t_1$, the resulting single pair
of values of $\digamma_{U235}(t_1)$ and $\digamma_{Pu239}(t_1)$ could
not be used to distinguish the two configurations. This is illustrated
in Fig.~\ref{fig:enrichment}, where the time evolution of the fission
fractions in uranium-235 and plutonium-239 is shown for both the NU
and LEU cores. The pair of fission fractions $\digamma_{U235}(t_1)$
and $\digamma_{Pu239}(t_1)$ for the NU core is nearly identical to the
the pair $\digamma_{U235}(t_2)$ and $\digamma_{Pu239}(t_2)$ for the
LEU core. This identity is approximate since it would require two
slightly different values of $t_2$ for uranium-235 and plutonium-239
to achieve exact identity, as indicated by the width of the gray
vertical band. This effect is, however, too small to distinguish the
two configurations. The speed at which both configurations move along
this path is significantly different, therefore comparing the
differential burn-up $\digamma(t+\Delta t)-\digamma(t)$, shown as
thick vertical lines, for both configurations gives rise to a
measurable difference between the NU and LEU cores, shown as thick red
lines. Note, that the uranium-238 fission fraction does not contribute
to this distinction since it stays constant for both core
configurations and the plutonium-241 fission fraction is present only
at a very small, basically unmeasurable level. Applying DBA to the
case at hand we find a 90\% confidence level distinction between the
two configurations solely based on antineutrino measurements within
about 160 days.

In summary, we have shown that if antineutrino monitoring of the
Iranian IR-40 reactor were instituted, it could provide a complete
assessment of the reactor core in terms of burn-up and plutonium
content with a sensitivity exceeding standard IAEA verification
requirements while meeting the timeliness criterion of 90 days. This
information could be available in a timely manner and could be
obtained by placing a detector outside the reactor building. This
technique does not rely on a declaration of reactor power since the
power could be inferred from the antineutrino signal simultaneously
with the core state. In case the reactor is shutdown for extended
periods, monitoring antineutrino emissions from long-lived fission
products could make it possible to verify the presence of the spent
fuel inside the reactor core for up to several hundred days after the
shutdown. In combination, these techniques could allow a graceful and
timely recovery from a loss of the continuity of
knowledge. Furthermore, differential burn-up analysis could provide a
means to distinguish different fuel enrichment levels. Other
safeguards methods alone could not achieve this performance, and are
likely to be more intrusive and labor intensive.

Antineutrino monitoring would work as well for \emph{any} reactor from
a few megawatts thermal power to small modular reactors to large scale
commercial nuclear power reactors. Also, it can and should be combined
with existing monitoring techniques to enhance effectiveness against a
host of future possible developments.

While the results of theoretical analyses are promising, antineutrino
reactor monitoring still faces the need for crucial R\&D in terms of
background rejection and rugged detection systems as well as a precise
calibration of reactor antineutrino fluxes. Looking ahead, and noting
Iran's willingness to extend IAEA access into aspects of its nuclear
program that are not available in other states, Iran may itself wish
for the IAEA to include antineutrino monitoring in the safeguards
approach for the IR-40, providing a real-world opportunity for a full
scale demonstration to enhance the credibility of the global
non-proliferation system.

\acknowledgements

 We  would like to thank H.~Kippe for providing the
 detailed reactor model of the IR-40 as the starting point for our
 calculation. This work was supported by the U.S. Department of Energy
 under contract \protect{DE-SC0003915} and by a Global Issues
 Initiative grant by the Institute for Society, Culture and
 Environment at Virginia Tech.

\bibliographystyle{apsrev} \bibliography{references}

\end{document}